\documentclass[aps,pra,twocolumn,groupedaddress,showpacs]{revtex4}

\usepackage{color}
\usepackage{graphicx}
\usepackage{amsfonts}
\usepackage{amsmath, amsthm, amssymb}
\usepackage{verbatim}
\usepackage{dcolumn}
\usepackage{ulem}

\newcommand{\etal}{\mbox{\it et~al.}}
\newcommand{\datastrut}{\vrule height12pt depth4pt width0pt}

\newcommand{\second}{\ensuremath{\mathrm{s}}}

\newcommand{\ms}{\ensuremath{\mathrm{m} \second }}

\newcommand{\Hz}{\ensuremath{\mathrm{Hz}}}

\newcommand{\GHz}{\ensuremath{\mathrm{G}\Hz}}
\newcommand{\MHz}{\ensuremath{\mathrm{M}\Hz}}
\newcommand{\kHz}{\ensuremath{\mathrm{k}\Hz}}
\newcommand{\mHz}{\ensuremath{\mathrm{m}\Hz}}

\newcommand{\vd}{\ensuremath{\vec{d}}}

\newcommand{\Tesla}{\ensuremath{\mathrm{T}}}

\newcommand{\uT}{\ensuremath{\mu\Tesla}}
\newcommand{\nT}{\ensuremath{\mathrm{n}\Tesla}}

\newcommand{\muw}{\ensuremath{\mu\mathrm{w}}}
\newcommand{\wrf}{\ensuremath{\omega_{\muw}}}
\newcommand{\watom}{\ensuremath{\omega_{0}}}

\newcommand{\nurf}{\ensuremath{\nu_{\muw}}}

\newcommand{\Ohm}{\ensuremath{\Omega}}

\newcommand{\MOhm}{\ensuremath{\mathrm{M}\Ohm}}
\newcommand{\GOhm}{\ensuremath{\mathrm{G}\Ohm}}
\newcommand{\mps}{\ensuremath{\mathrm{m}/\second}}

\newcommand{\mbar}{\ensuremath{\mathrm{mbar}}}

\newcommand{\Watt}{\ensuremath{\mathrm{W}}}
\newcommand{\mWatt}{\ensuremath{\mathrm{mW}}}
\newcommand{\muWatt}{\ensuremath{\mu\Watt}}

\newcommand{\cm}{\ensuremath{\mathrm{c}\meter}}

\newcommand{\cmc}{\ensuremath{\cm^{3}}}
\newcommand{\mm}{\ensuremath{\mathrm{m}\meter}}
\newcommand{\nm}{\ensuremath{\mathrm{n}\meter}}

\newcommand{\meter}{\ensuremath{\mathrm{m}}}

\newcommand{\Volt}{\ensuremath{\mathrm{V}}}
\newcommand{\kVolt}{\ensuremath{\mathrm{k}\mathrm{V}}}

\newcommand{\Ampere}{\ensuremath{\mathrm{A}}}

\newcommand{\muAmpere}{\ensuremath{\mathrm{\mu}\Ampere}}

\newcommand{\Efield}{\ensuremath{\vec{\mathcal{E}}}}

\newcommand{\EfieldModVec}{\ensuremath{|\vec{\mathcal{E}}|}}
\newcommand{\EfieldMod}{\ensuremath{\mathcal{E}}}
\newcommand{\Bfield}{\ensuremath{\mathcal{B}}}
\newcommand{\Ham}{\ensuremath{\mathbb{H}}}
\newcommand{\ssp}{\ensuremath{\alpha_0^{(2)}}}

\newcommand{\tsp}{\ensuremath{\alpha_0^{(3)}}}
\newcommand{\tspFfour}{\ensuremath{\tsp(F{=}4)}}
\newcommand{\tspFthree}{\ensuremath{\tsp(F{=}3)}}
\newcommand{\ttp}{\ensuremath{\alpha_2^{(3)}}}
\newcommand{\ttpFfour}{\ensuremath{\ttp(F{=}4)}}

\newcommand{\nkp}{\ensuremath{\alpha_k^{(n)}}}
\newcommand{\Hhf}{\ensuremath{\Ham_{hf}}}
\newcommand{\HSt}{\ensuremath{\Ham_{St}}}
\newcommand{\aFc}{\ensuremath{a_{F\kern-0.13em c}}}
\newcommand{\HFc}{\ensuremath{\Ham_{F\kern-0.13em c}}}

\newcommand{\ks}{\ensuremath{\mathrm{k}_{s}}}
\newcommand{\kt}{\ensuremath{\mathrm{k}_{t}}}

\newcommand{\HzkVsq}{\ensuremath{\Hz/\mathrm{kV}^2}}
\newcommand{\HzkVcms}{\ensuremath{\Hz/(\mathrm{kV}/\mathrm{cm})^2}}

\newcommand{\HzperkVsquared}{\ensuremath{\Hz/(\mathrm{k\Volt})^{2}}}
\newcommand{\HzperkVolt}{\ensuremath{\Hz/\mathrm{k\Volt}}}

\newcommand{\degree}{\ensuremath{^\circ}}

\def\bra#1{\mathinner{\langle{#1}|}}
\def\ket#1{\mathinner{|{#1}\rangle}}
\def\braket#1{\mathinner{\langle{#1}\rangle}}

{\catcode`\|=\active
  \gdef\Braket#1{\left<\mathcode`\|"8000\let|\bravert {#1}\right>}}
\def\bravert{\egroup\,\vrule\,\bgroup}

\begin{document}


%
\title{Measurement of the scalar third-order electric polarizability
  of the Cs ground state using CPT-spectroscopy in Ramsey
  geometry}


\author{Jean-Luc Robyr}
\altaffiliation{SYRTE-Observatoire de Paris, 75014 Paris, France}
\email{Jean-Luc.Robyr@obspm.fr}
\author{Paul Knowles}
\altaffiliation{Rilkeplatz 8/9, A--1040, Vienna, Austria}
\author{Antoine Weis}
\affiliation{Fribourg Atomic Physics Group,
  Department of Physics,
  University of Fribourg,
  Switzerland}
\homepage[]{http://physics.unifr.ch/en/page/89/}

\date{\today}

\begin{abstract}

The AC Stark shift induced by blackbody radiation is a major source of
systematic uncertainty in present-day cesium microwave frequency
standards.
The shift is parametrized in terms of a third-order electric
polarizability \tsp{} that can be inferred from the static electric
field displacement of the clock transition resonance. 
In this paper, we report on an all-optical CPT pump-probe experiment
measuring the differential polarizability
$\Delta\tsp = \tspFfour - \tspFthree$ on a thermal Cs atomic beam,
from which we infer $\tspFfour = 2.023(6)_{stat}(9)_{syst}~\HzkVcms$,
which corresponds to a scalar Stark shift parameter $\ks =
-2.312(7)_{stat}(10)_{syst}~\HzkVcms$. 
The result agrees within two standard deviations with a recent
measurement in an atomic fountain, and rules out another recent result
obtained in a Cs vapor cell. 
\end{abstract}

%

\maketitle
%
%
\section{INTRODUCTION}
%
Blackbody radiation (BBR) displaces, via the AC-Stark shift, the Cs
microwave clock transition frequency.
This effect is a dominant limitation in the accuracy of present
microwave atomic frequency standards at the $10^{-16}$
level~\cite{Itano:1982:SHS,Wynands:2005:AFC}.
Corrections of the clock frequency's BBR shift rely on a precise and
accurate knowledge of the electric polarizabilities describing the
shift (a contemporary, although not exhaustive review of atomic
polarizabilities is given by Mitroy~\etal{} \cite{Mitroy:2010:TAA}).
The most accurate measurements of the relevant parameter, expressed as
a difference of polarizabilities $\Delta\tsp$ (defined below), have
been extracted from Stark shift measurements in DC electric fields.
The motivation for the work herein arose from the $6\sigma$ disaccord
between the two most recent measurements of $\Delta\tsp$:
$4.564(8)~\HzkVcms$~\cite{Rosenbusch:2007:BRS} and
$4.10(8)~\HzkVcms$~\cite{Godone:2005:SSM}.
The work of \cite{Rosenbusch:2007:BRS}, which built on the work of
\cite{Simon:1998:MSS}, measured the shift using a static electric
field applied to atoms in an atomic fountain clock, whereas the work
of \cite{Godone:2005:SSM} used a static field applied (externally) to
an atomic Cs vapor confined in a glass cell.
We note that the result~\cite{Rosenbusch:2007:BRS} is in agreement
with recent theoretical values
\cite{Rosenbusch:2009:ASS,Angstmann:2006:FSC,Beloy:2006:HAC}.

In this work, we present an alternative experimental approach for
measuring $\Delta\tsp$.
An atomic beam technique was adapted to an all-optical pump-probe
experiment using coherent population trapping (CPT) both to create a
$\Delta m_F{=}0$ hyperfine coherence in the Cs ground state and to
subsequently probe the coherence following its evolution in applied
static electric and magnetic fields.
The resulting Ramsey resonance data were analyzed in two complementary
ways, first by a Fourier decomposition method applied to scans of the
fringe pattern~\cite{Shirley:1997:VDC,DiDomenico:2011:FAR}, and second
by tracking the central fringe's zero crossing, both measured as a
function of the applied electric field's magnitude.
In the following we present the conventional parametrization of the
effect, develop the signal model and its analysis by Fourier
decomposition, introduce pertinent details of the apparatus, and give
the analysis and results, including a discussion on limiting
systematics.

\section{THEORY}
%
%
\subsection{The Stark shift}
%
The interaction of an atom with an applied electric field \Efield{} is
described by the Stark Hamiltonian $\HSt = -\vd\cdot\Efield$.
The energy shift $\Delta E_{F,m_F}$ of a magnetic hyperfine sublevel
$\ket{n^2S_{1/2},F,m_F}$ of an alkali atom ground state is
parametrized as $\Delta E_{F,m_F} = -\frac{1}{2}\alpha |\Efield|^{2}$,
where the polarizability $\alpha$ is calculated by perturbation theory
using the Hamiltonian $\Ham=\HSt+\Hhf$, \Hhf{} being the hyperfine
interaction Hamiltonian.
Transition energies between internal atomic levels will consequently
change in proportion to the difference of the involved states'
polarizabilities.
The polarizability $\alpha$ is traditionally broken down via series
expansion in both the perturbation order $n$ at which the component
contributes and the multipole order $k$ of its interaction, which,
following the notation and methods established in
\cite{Angel:1968:HSS,Ulzega:2006:TES,Ulzega:2006:RET,Ulzega:2007:RET},
we will denote as \nkp.

The sublevel energies of the Cs 6S$_{1/2}$ ground state of interest
here are affected only by the polarizabilities $\ssp$, $\tsp$ and
$\ttp$, where the (by far dominating) scalar second order
polarizability \ssp{} is independent of $F$ and $m_F$, and therefore
does not contribute to a differential energy shift of the states
coupled by the clock transition.
The third order scalar polarizability \tsp{} depends only on $F$,
while the third order tensor polarizability \ttp{} depends both on $F$
and $m_F$, so that both values affect the clock transition's DC Stark
shift.
The relevant $\alpha$'s of the $F{=}3$ state are expressible as
constants times the polarizabilities of the $F{=}4$ state, so that the
electric field induced frequency shift of the $\Delta m_F{=}0$
transitions $\ket{6S_{1/2}, F{=}3,m_F}\rightarrow\ket{6S_{1/2},
  F{=}4,m_F}$ can be expressed as
\begin{subequations}
\begin{align}
\Delta\nu_\mathrm{Stark}({m_F})
=& -\frac{1}{2}\left[
           \frac{16}{7}\,\tsp{+}
           \frac{3 m_{F}^{2}{-}16}{28} f(\theta)\,\ttp\right]\EfieldModVec^2
\label{eq:StarkEqu}\\
\equiv& -\frac{1}{2}\,\alpha(m_F)\,\EfieldMod^2
\label{eq:StarkEqu2}\,,
\end{align}
\end{subequations}
where $\tsp{=}\tsp(F{=}4)$ and $\ttp{=}\ttp(F{=}4)$ are the
third-order scalar and tensor polarizabilities, respectively, of the
$F{=}4$ hyperfine state, $m_{F}$ is the magnetic quantum number
defined by the quantization axis (chosen along the magnetic field
$\hat{\Bfield}$), and $f(\theta){=}3\cos^{2}(\theta)-1$ with
$\cos\theta=\hat{\EfieldMod}\cdot\hat{\Bfield}$.
We note that a sign error in the evaluation of the \ttp{} term in the
original paper by~\cite{Sandars:1967:DPG} was corrected
in~\cite{Ulzega:2006:RET,Ulzega:2007:RET}.
Note also that we set $h{=}1$ in the definition of polarizabilities,
so that the latter are expressed in the practical `laboratory units'
of Hz/(kV/cm)$^2$.

The Stark shift, $\Delta\nu_\mathrm{Stark}({m_F{=}0})$, of the clock
transition $\ket{6S_{1/2}, F{=}3,m_F{=}0}{\rightarrow}\ket{6S_{1/2},
F{=}4,m_F{=}0}$ can be parametrized as
\begin{align}
\Delta\nu_\mathrm{Stark}({m_F{=}0})=\left[\ks+\kt\,f(\theta)\right]\,\EfieldModVec^2
\label{eq:ksktDef0}
\end{align}
with scalar and tensor constants, $\ks$ and $\kt$ that are related
to the quantities introduced above via
\begin{align}
\ks + \kt\,f(\theta)
&= -\frac{8}{7}\,\tsp+\frac{2}{7}\,\ttp\,f(\theta)\,.
\label{eq:ksktDef}
\end{align}

The clock shift arising from the blackbody spectrum is isotropic, implying
$\braket{f(\theta)}=0$, and is thus not sensitive to \ttp, i.e., to $k_t$.
Laboratory experiments, on the other hand, measuring the clock shift
by applied AC or DC fields must therefore consider \ttp{} (or $k_t$)
when extracting \tsp{} (or $k_s$) from the measured effect, while
model calculations of the BBS determine the scalar contribution,
\ks~\cite{Beloy:2006:HAC,Angstmann:2006:FSC}, only.

The dependence of the BBR shift on temperature $T$ is normally
\cite{Itano:1982:SHS} expressed in one of three ways
\begin{align}
\delta \nu(T) & = \frac{-\Delta\tsp}{2}
                  \left(\!831.9\,\frac{\mathrm{V}}{\mathrm{m}}\!\right)^2\!
                  \left(\!\frac{T}{300\:\mathrm{K}}\!\right)^4
\left[ 1 {+} \epsilon \left(\!\frac{T}{300\:\mathrm{K}}\!\right)^2\!\right],
\nonumber\\
 & = \ks
                  \left(\!831.9\,\frac{\mathrm{V}}{\mathrm{m}}\right)^2\!
                  \left(\!\frac{T}{300\:\mathrm{K}}\!\right)^4
\left[ 1 {+} \epsilon \left(\!\frac{T}{300\:\mathrm{K}}\!\right)^2\right],
\label{eq:tdep} \\
 & = \beta\,\nu_{00} \,\left(\!\frac{T}{300\:\mathrm{K}}\!\right)^4
\left[ 1 {+} \epsilon  \left(\!\frac{T}{300\:\mathrm{K}}\!\right)^2\right]\,,
\nonumber
\end{align}
where $\nu_{00}=9\,192\,631\,770$~Hz is the Cs clock transition frequency.
The correction factor $\epsilon$ was evaluated to be
0.014~\cite{Itano:1982:SHS} or 0.013~\cite{Angstmann:2006:FSH}.
Various publications present either \tsp(F),
$\Delta\tsp{}\left({\equiv}\tsp(F{=}4)-\tsp(F{=}3){=}+\dfrac{16}{7}\tspFfour\right)$,
$\ks\left({=}-\dfrac{8}{7}\tspFfour\right)$, 
or $\beta$.

\subsection{Experimental approach}
\begin{figure}[t]
\includegraphics[width=0.45\textwidth]{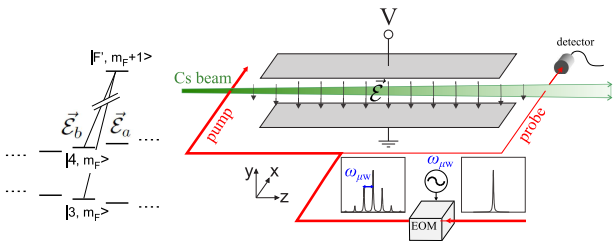}
\caption{(Color Online)
Principle of the all optical Ramsey method for measuring the Stark shift.
The carrier and a sideband (spaced by $\omega_{\muw}$) of a
phase-modulated laser beam are used to create a coherent superposition
of ground state hyperfine levels in a Cs atomic beam.
The phase accumulated by the coherence by the atoms' interaction with
combined electric and magnetic fields is probed by a weaker
bichromatic field whose microwave modulation has a fixed, but tunable
phase with respect to the pump field modulation. 
}
\label{RamseyInAtomicBeam_sketch}
\end{figure}

%
Experiments were carried out in an effusive cesium beam using an
all-optical Ramsey pump-probe technique relying on coherent population
trapping (CPT) by a bichromatic laser field, rather than on the
conventional Ramsey resonance method using (spatially or temporally)
separated interactions in microwave cavities, such as deployed in
atomic beams \cite{Carrico:1968:ABR,Gould:1969:QSS} or in fountain
clocks \cite{Rosenbusch:2007:BRS}.
Details of our method were already described in the literature
\cite{Robyr:2010:SSC,Robyr:2011:CPP,Robyr:2011:MTO}.
In brief, CPT pumping by a phase-coherent bichromatic optical field
with components of identical (circular or linear) polarization, and
with frequency splitting given by the microwave frequency \wrf{} is
used to create a specific coherent superposition of hyperfine-Zeeman
states $\ket{F{=}3,m_F}$ and $\ket{F{=}4,m_F}$ in the Cs ground state
(Fig.~\ref{RamseyInAtomicBeam_sketch}).
In a subsequent light-free evolution zone, the coherence oscillates at the
specific level splitting frequency \watom{} that is subject to
tunable static electric and magnetic fields.
In a probe zone, the phase accumulated by the hyperfine coherence
is compared to the phase accumulated by the constant evolution
of the transition-driving microwave oscillator.
When \wrf{} is scanned, Ramsey resonance signals are observed in the
transmitted probe laser power.
When a magnetic field lifts the Zeeman degeneracy, the full spectrum
contains six or seven distinct $\Delta m_F{=}0$ resonances depending
on the relative orientation of the magnetic and electric fields and
the light fields' polarization.

We have implemented two different methods of data collection and analysis.
In the first method, full Ramsey resonance curves were measured by
scanning \wrf{} for different electric field values, and the data were
reduced by Fourier decomposition.

In the second method, active feedback was used to lock the frequency
\wrf{} of the microwave generator (producing the bichromatic laser
field by phase modulation) to the zero crossing of the central
dispersive Ramsey fringe, allowing a tracking of \wrf{} as a function
of the applied electric field strength.

\subsection{Atomic pump-evolution-probe model}
%
Consider the two hyperfine ground states
$\ket{a}{=}\ket{6S_{1/2},F{=}3,m_F}$ and
$\ket{b}{=}\ket{6S_{1/2},F{=}4,m_F}$ (inset of
Fig.~\ref{RamseyInAtomicBeam_sketch}) whose energies differ by the
hyperfine splitting energy $\hbar\watom$.
The bichromatic pump light field
\begin{align}
\vec{\mathcal{E}}_\mathrm{pump}(t)&= \vec{\mathcal{E}}_a\,e^{-i\omega_a t}   +
\vec{\mathcal{E}}_b\,e^{-i\omega_b t}\\
&=\vec{\mathcal{E}}_a\,e^{-i\omega t}   +
\vec{\mathcal{E}}_b\,e^{-i(\omega-\wrf) t}\,,
\end{align}
where the fields $\vec{\mathcal{E}}_a$ and $\vec{\mathcal{E}}_b$ are
near resonant with transitions $\ket{a}\rightarrow\ket{e}$ and
$\ket{b}\rightarrow\ket{e}$ to a specific excited state
$\ket{e}=\ket{6P_{1/2},F',m'_F}$  will put the atoms into a coherent
superposition state 
\begin{align}
\ket{\Psi_0}&=\eta_a\,\ket{a}+\eta_b\ket{b}\,e^{-i\watom t}\,, \label{eq:PsiIni}
\end{align}
where we have omitted the (irrelevant) phase factor $e^{-i\omega t}$.
When probed by a weaker bichromatic field
\begin{align}
\vec{\mathcal{E}}_\mathrm{probe}(t)&= \varepsilon\,\vec{\mathcal{E}}_a\,e^{-i\omega t}   +
\varepsilon\, \vec{\mathcal{E}}_b\,e^{-i(\omega-\wrf) t}\,,
\end{align}
that is in-phase with the pump field,  the power absorbed from the
probe beam is given by 
\begin{align}
\Delta P_\mathrm{probe}&\propto\varepsilon^2
\left|\bra{e}\vec{d}\cdot\vec{\mathcal{E}}_\mathrm{probe}\ket{\Psi_0}\right|^2\\ 
&\propto\left|\eta_a\,\vec{d}_{ae}\cdot\vec{\mathcal{E}}_a +
\eta_b\,\vec{d}_{be}\cdot\vec{\mathcal{E}}_b\,e^{i(\wrf-\watom)t}\right|^2\\ 
&\propto\,\left|\eta_a\,\Omega_{ae} +
\eta_b\,\Omega_{be}\,e^{i(\wrf-\watom)t}\right|^2\,, 
\end{align}
where $\vec{d}_{ie}$ denotes the electric dipole matrix elements
$\bra{e}\vec{d}\ket{i}$, and where the Rabi frequencies are defined by
$\hbar\Omega_{ie}=\vec{d}_{ie}\cdot\vec{\mathcal{E}}_i$. 
On resonance, $\wrf=\watom$, the absorbed power will vanish, when the
state amplitudes $\eta_{a,b}$ obey 
\begin{equation}\label{eq:DarkStateCondition}
 \frac{\eta_a}{\eta_b}=-\frac{\Omega_{be}}{\Omega_{ae}}
   = -\frac{\vec{d}_{be}\cdot\vec{\mathcal{E}}_b\,}{\vec{d}_{ae}\cdot\vec{\mathcal{E}}_a}\,, 
\end{equation}
and the corresponding (normalized) state
\begin{align}
\ket{\Psi_0}&=\frac{\Omega_{be}}{\sqrt{\Omega_{ae}^2+\Omega_{be}^2}}\,\ket{a}-
\frac{\Omega_{ae}}{\sqrt{\Omega_{ae}^2+\Omega_{be}^2}}\ket{b}
 \,e^{-i\watom t} 
\label{eq:PsiIni2} 
\end{align}
is called a dark state, since it does not absorb light, and hence does
not emit fluorescence radiation.
The formation of dark states, their perturbation by interactions with
external fields, and their detection via light absorption (or
fluorescence) forms the basis of coherent population trapping (CPT)
spectroscopy.
With the matrix elements being given, the amplitudes
$\mathcal{E}_{a,b}$ of the two field components can be chosen such
that the state amplitudes in Eq.~\eqref{eq:DarkStateCondition} obey
$\eta_a=-\eta_b$, so that the power absorbed from the probe beam
interrogating the atoms after an evolution time $T$ takes on the
simple form
\begin{align}
\Delta P_\mathrm{probe}&
      \propto \left|1{-}e^{i(\wrf{-}\watom)T}\right|^2 
      \propto 1{-}\cos\left[(\watom{-}\wrf)\,T\right]\,.
\label{eq:DeltaP2}
\end{align}

\subsection{Phase and frequency shifts of the Ramsey spectrum}
%
The expressions above have assumed that the bichromatic fields in
the pump and probe regions oscillate in phase.
Since the pump and probe beams travel different paths from the
source to their respective interaction zones, the relative phase
between their two frequency components will acquire an additional
spatial phase shift
\begin{align}
\Delta\varphi_\mathrm{path} = \wrf \, \frac{\Delta x}{c}\,,
\label{eq:phipath}
\end{align}
where $\Delta x$ is the difference of the paths travelled by the pump
and probe beams.
This phase propagates through the calculation and modifies the
detuning-dependent term in Eq.~\eqref{eq:DeltaP2} to
\begin{align}
\Delta P_\mathrm{probe}  \propto
\cos\left[(\watom -\wrf)\,T-\Delta\varphi_\mathrm{path}\right]\,.
\label{eq:2kappa8}
\end{align}

Choosing the propagation phase to be
$\Delta\varphi_\mathrm{path} = 0\mod\pi$ will yield symmetric
(absorptive) Ramsey fringes with respect to the line center
($\wrf{=}\watom$), while $\Delta\varphi_\mathrm{path} = \pi/2\mod\pi$
will produce antisymmetric (dispersive) fringes.
We define the Ramsey signal $S(\wrf)$ as being the velocity averaged
change of the probe transmission \eqref{eq:2kappa8}, viz.,
\begin{widetext}
\begin{align}
S(\wrf)=\int\limits_{0}^{\infty}\rho(v)\,\cos\left\{
\left[\watom+\Delta\omega(m_F)-\wrf\right]
  \frac{L}{v}-\Delta\varphi(m_F)\right\}\,dv\;,
\label{eq:RamseyFormula}
\end{align}
\end{widetext}
where $\rho(v)$ is the atomic velocity distribution, and where the
Stark and Zeeman shifts, i.e., the frequency shift induced by the
electric and magnetic fields, respectively, are given by
\begin{equation}
\Delta\omega(m_F)=\Delta\omega_\mathrm{Stark}(m_F)+\Delta\omega_\mathrm{Zeeman}(m_F)\,,
\end{equation}
while
\begin{equation}
\Delta\varphi(m_F)=\Delta\varphi_\mathrm{path}(m_F)+\Delta\varphi_\mathrm{mot}(m_F)
\end{equation}
represents the sum of phase shifts due to the pump and probe beams
path length differences and due to the motional Zeeman effects,
respectively, as discussed in more detail below.
For hyperfine coherences formed by the pair of states $\ket{3, m_F}$
and $\ket{4, m_F}$, the (differential) frequency shift induced by the
Stark interaction of interest is
$\Delta\omega_\mathrm{Stark}{=}2\pi\,\Delta \nu_\mathrm{Stark}(m_F)$,
where $\Delta \nu_\mathrm{Stark}(m_F)$ is given by
Eq.~\eqref{eq:StarkEqu}.
The Zeeman frequency shift induced by the static applied magnetic
field $\vec{\mathcal{B}}$ is given by $\Delta\omega_\mathrm{Zeeman}=
(\gamma_4{-}\gamma_3)\,|\vec{\mathcal{B}}|\,m_F$, where the $\gamma_F$
are the gyromagnetic ratios of the hyperfine levels $F$.

In a monochromatic beam, all atoms will have the same time of flight
$T$ between the pump and probe zones, leading to a cosine-like
oscillatory dependence \eqref{eq:2kappa8} on the microwave detuning.
Averaging over the broad velocity distribution of the thermal beam
used in the experiment will impose the typical Ramsey-type of envelope
\eqref{eq:RamseyFormula} on that oscillatory structure.
The velocity distribution $\rho(v)$ is not known \textit{a priori}
since it differs appreciably from a perfect Maxwell-Boltzmann
distribution in our effusive beam, the discrepancy arising from atomic
collisions in the oven's nozzle and collimation regions
\cite{Ramsey:1956:MB}.

\subsection{Fourier transform of the Ramsey spectrum}
\label{sec:FourierTheory}
%
The real and imaginary parts of the (inverse) Fourier transform of
Eq.~\eqref{eq:RamseyFormula} are given by
\begin{align}
A(t)&\equiv\mathrm{Re}[\mathcal{F}^{-1}[S(\wrf)]]\\
&= \sqrt{\frac{\pi}{2}}\rho(t)
\frac{L}{t^{2}}\cos\left[(\watom+\Delta\omega)
t-\Delta\varphi\right]\,,
\end{align}
and
\begin{align}
B(t)&\equiv\mathrm{Im}[\mathcal{F}^{-1}[S(\wrf)]]\\
&= \sqrt{\frac{\pi}{2}}\rho(t)
\frac{L}{t^{2}}\sin\left[(\watom+\Delta\omega)
t-\Delta\varphi\right]\,,
\end{align}
where $\Delta\omega$ represents the sum of the Zeeman and Stark
shifts in \eqref{eq:RamseyFormula}.
These Fourier transforms allow the determination of the time of
flight distribution
\begin{align}
\rho(t) = &\frac{t^{2}}{L} \sqrt{\frac{2}{\pi}}
           \sqrt{A^{2}(t) +B^{2}(t)}
\label{equ:numberdensity}
\end{align}
between the pump and probe zones (separated by $L$), and the phase
\begin{align}
(\watom
+\Delta\omega)\,t-\Delta\varphi=&\arctan\left(\frac{B(t)}{A(t)}\right).
\label{equ:fourierphase}
\end{align}
accumulated by the atoms on their flight from the pump to the probe
zone.
%
\subsection{Effects of a static electric field on the Ramsey pattern}
\label{sec:FieldEffects}
%
Subjecting the atoms to an electric field during their flight between
the pump and probe interactions changes the energies of the hyperfine
sublevels both via the Stark interaction proper and via the
Zeeman effect induced by the motional magnetic field experienced by
the atoms moving through the transverse electric field.
Both effects appear simultaneously and modify the Ramsey lineshape
of Eq.~\eqref{eq:RamseyFormula} in different ways, due to their
different velocity dependencies.

\begin{figure}[t]
\includegraphics[width=0.45\textwidth]{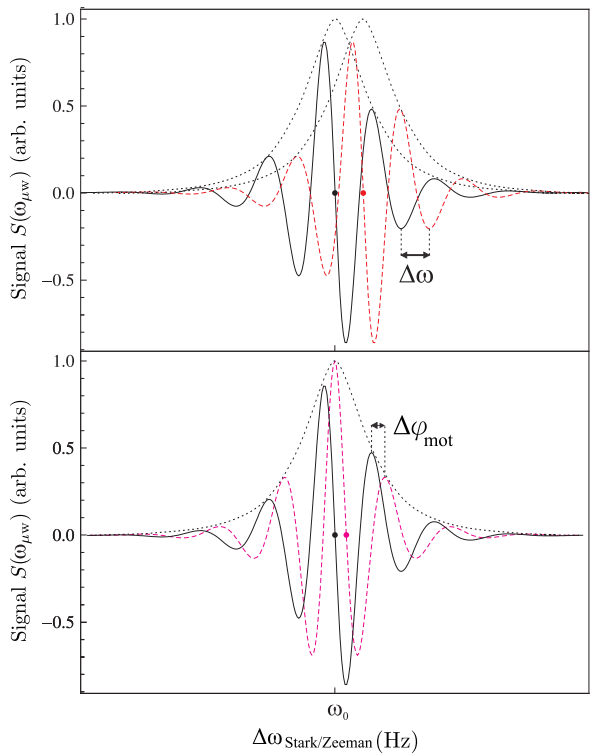}
\caption{(Color Online)
  Expected unperturbed dispersive Ramsey lineshape for
  $\Delta\varphi_\mathrm{path}{=}\pi/2$ (black line), and its change
  (red, dashed line) by a frequency shift
  $\Delta\omega_\mathrm{Stark/Zeeman}$ (top) and the motional phase
  shift $\Delta\varphi_\mathrm{mot}$ (bottom). 
  The frequency shift displaces both the fringes and their envelope,
  while the phase shift displaces the fringes under the envelope.
  Both effects therefore displace the zero crossings (marked by dots).}
\label{FringesPHandFREQchanges}
\end{figure}

The frequency shift $\Delta\omega_\mathrm{Stark}$ that the $\Delta
m_F{=}0$ coherence acquires in the free evolution zone $L$ due to the
direct Stark interaction with an electric field of spatial
distribution $\Efield(z)$ is given by
\begin{subequations}
\begin{align}
\Delta\omega_\mathrm{Stark}&=-\frac{1}{2}\,\alpha(m_F)\,
\langle|\Efield|^2\rangle_L
\label{eq:Stark_phase_a}\\ 
&=-\frac{1}{2}\,\alpha(m_F)\,\frac{1}{L}
\int\limits^{L}_{0}|\Efield(z)|^{2}\,dz
\label{eq:Stark_phase_b}\,,
\end{align}
\end{subequations}
where $\alpha(m_F)$ was introduced in Eq.~\eqref{eq:StarkEqu2}.
The Stark shift proper will thus shift the entire Ramsey spectrum
(fringes plus envelope) along the \wrf{} axis as shown on the top of
Fig.~\ref{FringesPHandFREQchanges}.

The electric field has a second effect on the Ramsey spectrum.
In the reference frame of atoms moving at velocity $\vec{v}$ through
the (transverse) electric field appears a motional magnetic field that
is given, to first order in $v/c$, by
\begin{align}
\vec{\Bfield}_{\mathrm{mot}}=\frac{\vec{v}}{c^2}\times\Efield\,.
\end{align}
For a typical mean atomic velocity $\overline{v}$ of $\sim 250~\mps$
and an electric field strength $|\Efield|$ of $\sim 20$~kV/cm the
motional field is $|\vec{\Bfield}_{\mathrm{mot}}|\sim 5~\nT$ for
$\Efield\perp\vec{v}$.
Atoms with a $\Delta m_{F}{=}0$ coherence will thus acquire, via the
linear Zeeman effect induced by $\vec{\mathcal{B}}_\mathrm{mot}$, a
phase shift
\begin{subequations}\label{eq:MotionalPhaseShift}
\begin{align}
\Delta\varphi_{\mathrm{mot}}
=&\frac{\gamma_4-\gamma_3}{c^{2}}\,m_{F}\,\sin(\xi)\,v\int\limits^{L}_{0}|\Efield(z)|\,dt\label{eq:MotionalPhaseShift_a}\\
=&\frac{\gamma_4-\gamma_3}{c^{2}}\,m_{F}\,\sin(\xi)\int\limits^{L}_{0}|\Efield(z)|\,dz\,,
\label{eq:MotionalPhaseShift_b}
\end{align}
\end{subequations}
where $\xi$ is the angle between $\vec{v}$ and $\Efield$.
In contrast to the frequency shift induced by the Stark interaction of
interest, the motional field effect is a topological (velocity
independent) phase shift that manifests itself as a displacement of
the Ramsey fringes under their otherwise fixed envelope, as shown in
the bottom graph of Fig.~\ref{FringesPHandFREQchanges}
We note that both effects displace the zero crossings of the Ramsey fringes.

\section{EXPERIMENTAL APPARATUS}

The main elements of the apparatus have been presented
in~\cite{Robyr:2010:SSC,Robyr:2011:CPP,Robyr:2011:MTO}.
We have used two methods for measuring the differential Stark shift of
the Cs clock transition.
Some elements of the experimental method used to lock the microwave
frequency to the Ramsey fringe center were presented
in~\cite{Robyr:2009:SSC} and a preliminary data analysis using the
Fourier transform method was reported
in~\cite{Robyr:2011:MTO,Robyr:2011:CPP}.
Here we presents the final analysis and results of date obtained with
the fringe tracking and Fourier transform methods, respectively.

\begin{figure}[tp]
\centerline{\includegraphics*[width=0.45\textwidth]{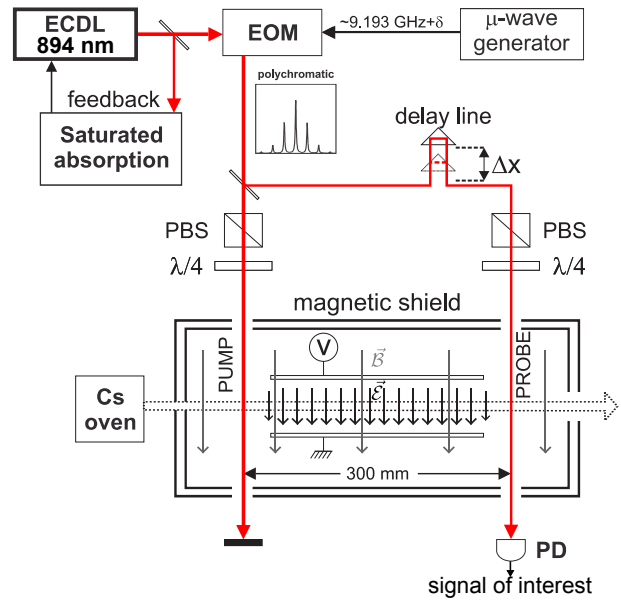}}
\caption{(Color Online) General experimental configuration of the
  Ramsey interrogation scheme. Details are given in the text.} 
\label{fig:exp_scheme_general}
\end{figure}

\subsection{Atomic beam}

A thermal Cs atomic beam is produced
by an effusive reflux oven inspired by the design
described in~\cite{Swenumson:1981:CFR}.
Additional collimating elements are placed along the atomic beam
trajectory to restrict the beam's cross section (horizontal and
vertical divergence angles of $\pm~4.5$ and $\pm~8.5$~mrad,
respectively) to a rectangular profile of $9~\mm$ along $\hat{x}$ and
$4~\mm$ along $\hat{y}$ in the Ramsey interrogation zone.
The atomic densities in the pump and probe interaction regions are
estimated to be $\sim 50 \times 10^{6}/\cmc$ and
$\sim 16 \times 10^{6}/\cmc$, respectively.
The pump, probe, and electric field regions are enclosed in a vacuum
chamber with typical pressure of $5\times 10^{-7}~\mbar$ surrounded by
a cylindrical two-layer $\mu$-metal shield.
A nominally homogeneous transverse magnetic field of $3.57(1)~\uT$ is
applied to all regions of interest.
The laser beams enter and leave the vacuum chamber through 10~\mm{}
diameter windows.

\subsection{Electric field generation and calibration}
\label{sec:Electric field and Calibration}

\begin{figure}[b]
\vspace{3mm}
\centering\includegraphics[width=0.99\linewidth]{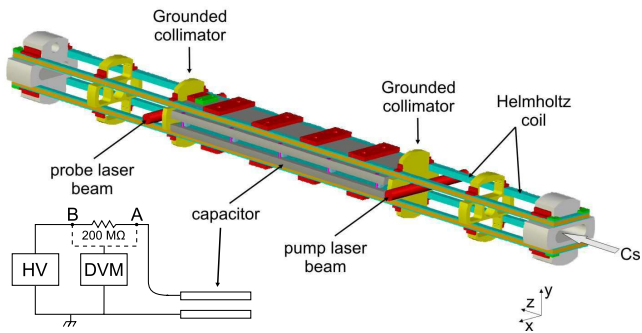}
\caption{(Color Online) View of the electrode support structure.
Grounded metallic collimator plates placed upstream and downstream of
the capacitor prevent the fringe fields from perturbing the pump and
probe interaction zones.
Long rectangular Helmholtz coils produce a homogeneous DC magnetic field
parallel to the electric field over the entire Ramsey interaction
region.
Inset shows the electrical connection scheme.
}
\label{Electrode_Support}
\end{figure}

A complete view of the field generating capacitor and its dielectric
support structure is shown in Fig.~\ref{Electrode_Support}.
The electrodes are made of two rectangular ($50~\mm \times 260 ~\mm
\times 4 ~\mm$) float glass plates with a conductive coating.
The plate spacing is defined by 10 optical flats with a thickness of
$6.065(1)~\mm$, inserted in a insulating polycarbonate holder.
Two grounded metal plates at each end of the capacitor are used to
collimate the Cs beam and to prevent the electric fringe fields from
perturbing the two optical interaction regions.

High voltage, provided by a Heinzinger PNC $60000$--$1$ump power
supply capable of delivering up to $60~\kVolt$ with a stability of
$10^{-4}$ over 8~hours is applied to one electrode while the other
electrode is grounded.
The maximum voltage used in the experiment was $10~\kVolt$ to avoid
sparking inside the vacuum chamber.
The electrode voltage is measured by a digital voltmeter
with a resolution of $10^{-4}$; the device was calibrated just before
the final measurements at the Swiss metrology institute METAS\@.
A~$200~\MOhm$ protection resistor between the generator and the high
voltage electrode is used to limit the current, and hence the
destructive power, of any sparks generated during breakdown.
For the Fourier transform measurements described below, the voltage
drop over the field-producing capacitor was measured directly (DVM
measuring at the capacitor plate, point `A'), while for the
fringe tracking experiments, we measured the voltage at the power
supply (DVM connected at point `B'), leading---because of leakage
currents through the resistor---to an increased uncertainty of the
electric field.

\begin{figure}[ht]
\centerline{\includegraphics[width=0.9\linewidth]{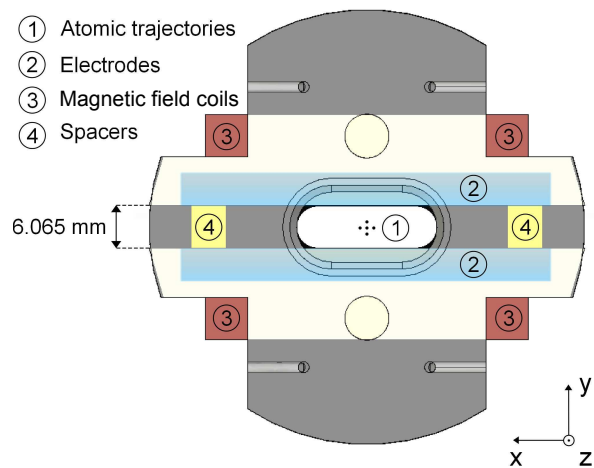}}
\caption{(Color Online) Inbound-atom view of the second collimator
  plate.
  This corresponds to a section view of the electrode support
  of Fig.~\ref{Electrode_Support} at the `capacitor' label position.
  The 5~dots at the center show the atom trajectory positions
  considered in the electric field modelling.}
\label{fig:At_Beam_trajec}
\end{figure}

In order to determine the average electric field applied to the atoms
from measurement of the voltage applied to the electrodes, the
electric field's spatial distribution was modelled by Dr.~Z.~Andjelic
from the ABB Corporate Research in Baden (CH), using the code
POLOPT~\cite{Andjelic:POL} that is based on the advanced boundary
integral method in three dimensions described
in~\cite{Andjelic:2008:BON}.
The modelling relied on the accurate representation of the mechanical
structures supporting the capacitor plates including all surfaces,
conductors and insulators in the region between the two grounded
collimation slits.
The model calculates the electric field throughout the volume for a
potential of $1~\Volt$ applied to one electrode, with the other held
at ground, and assumes that the field increases linearly with the applied
voltage.

The three vector components of \Efield{} are calculated every 1.75,
1.5, and 2~\mm{} along the $x$-, $y$-, and $z$-axis, respectively.
Since the modelling produces the electric field throughout all space
between the electrodes, the mean field values and their standard
deviations are evaluated for several possible atomic trajectories.
Along the $y$-axis, the trajectories that contribute to the signal are
limited by the probe laser beam's extension of $\pm1.2~\mm$.
Figure~\ref{fig:At_Beam_trajec} represents an ``atom's-eye'' view of
the apparatus, and shows the positions of the trajectories chosen for
averaging.

In order to relate the frequency shift $\Delta\omega_\mathrm{Stark}$
and the motional phase shift $\Delta\varphi_{mot}$ to the field
integrals in Eqs.~\eqref{eq:Stark_phase_b} and
\eqref{eq:MotionalPhaseShift_b}, respectively, one has to know
pump-probe separation $L$.
This distance was measured directly on the apparatus by determining
the separation of the centers of the pump and probe beams' intensity
distributions.
In order to account for a possible non-parallelism of the beams the
separations measured near the entrance and exit windows of the vacuum
chamber was averaged, yielding $L=301.8(7)~\mm$.

The $L$-averaged electric field integrals, defined as
\begin{subequations}\label{eq:I1I2Def}
\begin{align}
I_1&=\int\limits^{L}_{0}|\Efield(z)|\,dz\equiv
\frac{L}{d_\mathrm{eff}}\,(1\mathrm{V})
\label{eq:I1Def}\\
I_2&=\frac{1}{L}\int\limits^{L}_{0}|\Efield|^{2}(z)\,dz\equiv
\frac{1\mathrm{V}^2}{(d^2)_\mathrm{eff}}
\label{eq:I2Def}
\end{align}
\end{subequations}
for the five paths are given in Table~\ref{tab:EfieldCallib}, along
with the final average value.

\newcolumntype{k}[1]{D{.}{.}{#1}}
\begin{table}[ht]
\caption{Numerical values for the two electric field integrals defined
  by Eqs.~\eqref{eq:I1I2Def} obtained by field modelling, with a pump-probe
  separation $L=301.8(7)~\mm$ and a potential difference of $1~\Volt$.
  The uncertainty on the individual path
  calibration constants is dominated by the uncertainty on $L$,
  however, the uncertainty on the average is the standard deviation of
  the values.}
\centerline{
\begin{tabular}{ccccc} \hline\hline
   Trajectory $(x,y)$ & &{$I_1$ in V} & & {$I_2$\;in (V/cm)$^2$ }
   \datastrut\\ \hline \datastrut
    $(0,0)$       & &44.4208 & & 2.393(6) \\
    $(0,+1.5\mm)$ & &44.9346 & & 2.434(6) \\
    $(0,-1.5\mm)$ & &44.2066   & & 2.382(6)   \\
$(\pm 1.75\mm,0)$ & &44.4154   & & 2.392(6) \\ \hline
    Average       & &44.48(12) & & 2.399(9)\datastrut  \\ \hline\hline
\end{tabular}}
\label{tab:EfieldCallib}
\end{table}

Since the modelling calculations were done for a voltage of 1~V, the
integrals can be parametrized in terms of effective plate spacings
$d_\mathrm{eff}$ and $(d^2)_\mathrm{eff}$, defined by
Eqs.~\eqref{eq:I1I2Def}, that represent the effect of the capacitor's
finite size.
The numerical values of the effective spacings from the modelling
calculation are compared to the geometrical electrode spacing in
Table~\ref{tab:EfieldCallib2}.

\begin{table}[hb]
\caption{Comparison of electrode spacings.
Upper two values: Effective spacings from the modelled average value of
the electric field and its square as defined by
Eqs.~\eqref{eq:I1I2Def} with values from Table~\ref{tab:EfieldCallib}.
Bottom: Geometrical electrode spacing determined by spacers.
The difference between geometrical and effective separations is
dominated by the ratio of $L=301.8~\mm$, the pump-probe distance, to
the physical capacitor length of $260.0~\mm$ ($260.0~\mm /301.8~\mm
=0.86$, $d_\mathrm{geom} / d_\mathrm{eff} = 0.89$), with the remaining
difference due to the structure of the fringe fields.
}
\begin{tabular}{llcl}
 \hline\hline\datastrut
  $d_\mathrm{eff}$ &= $L/I_1$& &\phantom{6}6.79(2)    \phantom{11}mm \\
  $(d^2)_\mathrm{eff}$&= $1/I_2$& & 41.69(16) \phantom{1}mm$^2$\\
  $\sqrt{(d^2)_\mathrm{eff}}$& = $1/\sqrt{I_2}$& & \phantom{6}6.457(12) mm \\
  $d_\mathrm{geom}$ &&  & \phantom{6}6.065(1) \phantom{1}mm \\
 \hline\hline
\end{tabular}
\label{tab:EfieldCallib2}
\end{table}

\subsection{The bichromatic laser fields}
\label{sec:laser} 

The (single) laser used for the pump-probe experiments is a
$40~\mWatt$  extended cavity diode
laser emitting monochromatic radiation near the Cs $D_{1}$
transitions ($\sim 894.6~\nm$) with a spectral linewidth below
$1~\MHz$.
The laser wavelength is actively stabilized to the $F{=}3\rightarrow
F'{=}3 $ hyperfine component of the $D_{1}$ line using Doppler-free
spectroscopy in an auxiliary Cs vapor cell.

The phase-coherent bichromatic light field needed for the CPT
pump-probe scheme by a polarization maintaining fiber coupled
lithium niobate electro-optic phase modulator (Photline, model
NIR-MPX800-LN08).
The EOM is driven by a frequency-tunable microwave source (Rohde \&
Schwarz, model SMP 02 signal generator) capable of delivering
frequencies from $10~\MHz$ to $20~\GHz$ with a resolution of
$0.1~\Hz$, but here used to create $\nurf$ close to the Cs hyperfine
frequency of $9.2~\GHz$.
The microwave generator is referenced to the 10~MHz signal from an
atomic clock (Temex, model PFRS  Rb clock) for better stability and
accuracy.
The EOM creates sidebands at positive and negative integer multiples
of $\nurf$ around the laser carrier frequency.
The carrier and first sideband form the two components of the
phase-coherent bichromatic field.
The modulation index is chosen to yield  identical Rabi frequencies
of the two transitions ($6S_{1/2}, F'{=}3 \rightarrow 6P_{1/2}, F{=}3$
and $6S_{1/2}, F{=}4 \rightarrow 6P_{1/2}, F'{=}3$), whose oscillator
strengths are in the ratio $1:3$, and that are driven coherently by
the bichromatic field.

A $\sim 500~\muWatt$ pump beam and a $\sim 10~\muWatt$ probe beam are
derived from the modulated beam using a beam splitter as schematized
in Fig.~\ref{fig:exp_scheme_general}.
The probe beam path length and hence the spatial phase factor given by
Eq.~\eqref{eq:2kappa8} is varied by a delay line formed by two
orthogonal mirrors on a micrometer-controlled linear translation
stage.
Each change of the optical path length difference $\Delta L$ by one
microwave wavelength ($\lambda_{\muw}\cong 32.6~\mm$ for
$\nurf{=}9.192~\GHz$) changes the phase of the Ramsey pattern by
$2\pi$.
Phase adjustments were made with no electric field applied to the
atoms.
%
\subsection{Heterodyne detection}
\label{sec:heterodyne}
\begin{figure}[tp]
\centerline{\includegraphics*[width=0.45\textwidth]{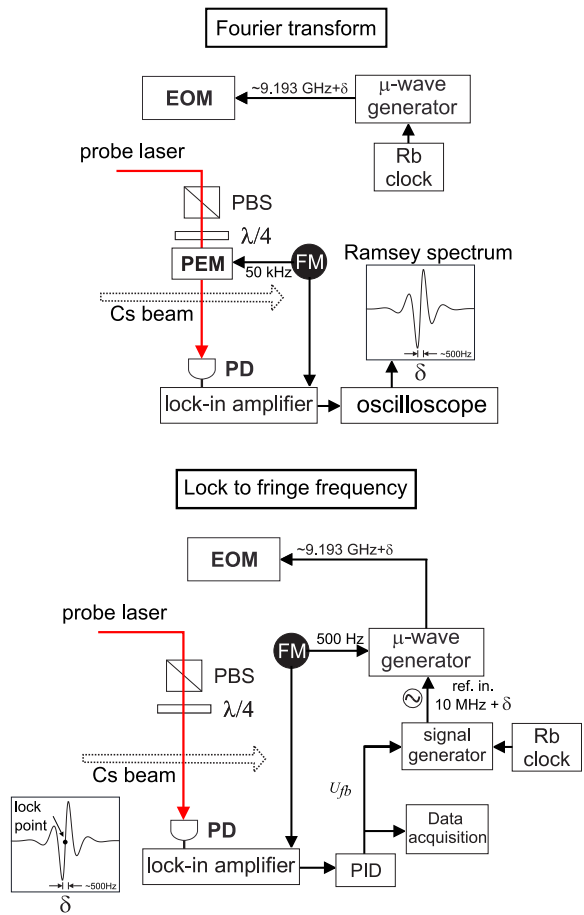}}
\caption{(Color Online)
Experimental schemes of the two heterodyne detection methods deployed here.
Top: Heterodyne detection used to record Ramsey spectra for Fourier
transform analysis.
The polarization of the probe beam is modulated between $\sigma^{+}$
and $\sigma^{-}$ with a photoelastic modulator at $50~\kHz$.
Bottom: The slope of the dispersive Ramsey fringe's zero crossing is used as
discriminant to lock the microwave frequency to the fringe center.
The heterodyne measurement is done by modulating the microwave
generator frequency, and the error signal, after PID-amplification, is
sent to the frequency modulation input of the $10~\MHz$ reference of
the microwave generator.
}
\label{fig:exp_scheme_heterodyne}
\end{figure}

%
Two distinct heterodyne detection methods, shown in
Fig.~\ref{fig:exp_scheme_heterodyne}, were used to measure the shift
of the central Ramsey fringe's frequency as a function of the
applied electric field applied to the atoms.

The first method consists in recording complete Ramsey spectra by
monitoring the transmitted probe laser power while scanning the
microwave frequency \wrf{} over a given $\ket{3,
m_F}\rightarrow\ket{4, m_F}$ transition.
In these experiments the pump beam components were
$\sigma^+$-polarized and the field configuration was $\Efield \perp
\vec{\Bfield}$, corresponding to $\theta=\pi/2$ in
Eq.~\eqref{eq:StarkEqu}.
In the probe region we measured the circular dichroism of the beam
by switching the probe beam polarization between $\sigma^+$ and
$\sigma^-$ states using a photoelastic modulator (PEM, Hinds
Instruments, model I/FS50).
The transimpedance-amplified (400~\kHz{} bandwidth) photodiode
signal was analyzed by a lock-in amplifier referenced to the
modulation frequency ($50~\kHz$) of the PEM.

In the second method the microwave frequency was actively locked to
the central Ramsey fringe's zero crossing and the electric field
induced frequency shift of the resonance frequency was inferred from
the feedback signal.
For these experiments the pump-probe path length difference $\Delta
x$ was set to a multiple of $\lambda_{\wrf}$, thus yielding
cosine-like Ramsey fringes (the absorptive counterpart of the
spectra shown in Fig.~\ref{FringesPHandFREQchanges}).
The frequency of the microwave oscillator was modulated by
23~kHz with a frequency of 525~Hz and the probe
detector's photocurrent demodulated by a lock-in amplifier locked to
that modulation frequency.
Scanning the microwave frequency \wrf{} then yields a dispersive
fringe pattern (the derivative of the absorptive cosine-pattern)
similar to the one shown in Fig.~\ref{FringesPHandFREQchanges}.
The near-resonance linear zero crossing of this signal is used as a
discriminator signal to form, after PID amplification, a feedback
signal $U_\mathrm{fb}$ controlling the frequency of the microwave
oscillator.
This control was achieved in the following way:
The Rb clock references a function generator (Agilent, model 33220A), which
generates the $10~\MHz$ reference signal for the microwave generator.
In order to lock the microwave frequency to the atomic signal, the
$10~\MHz$ generated by the function generator is controlled by
applying  $U_{fb}$ to the generator's FM control input.
The feedback signal $U_{fb}$ in the locked mode is recorded on a
digital oscilloscope for $\sim$600~$\second$ and its average value is
determined as a function of the applied electrode voltage $U$.
The calibration constant converting the feedback voltage into
frequency units is used to express the displacement of the Ramsey
fringe center in $\Hz$.
The pump and probe laser beams were $\sigma^+$-polarized for
experiments with $\Efield\perp \vec{\mathcal{B}}$ and linearly polarized for
$\Efield\parallel \vec{\mathcal{B}}$.

\section{Measurements}
%
\subsection{CPT-Ramsey spectra with $\mathcal{E}=0$}

\begin{figure}[tb!]
\centering

\includegraphics[width=0.4\textwidth]{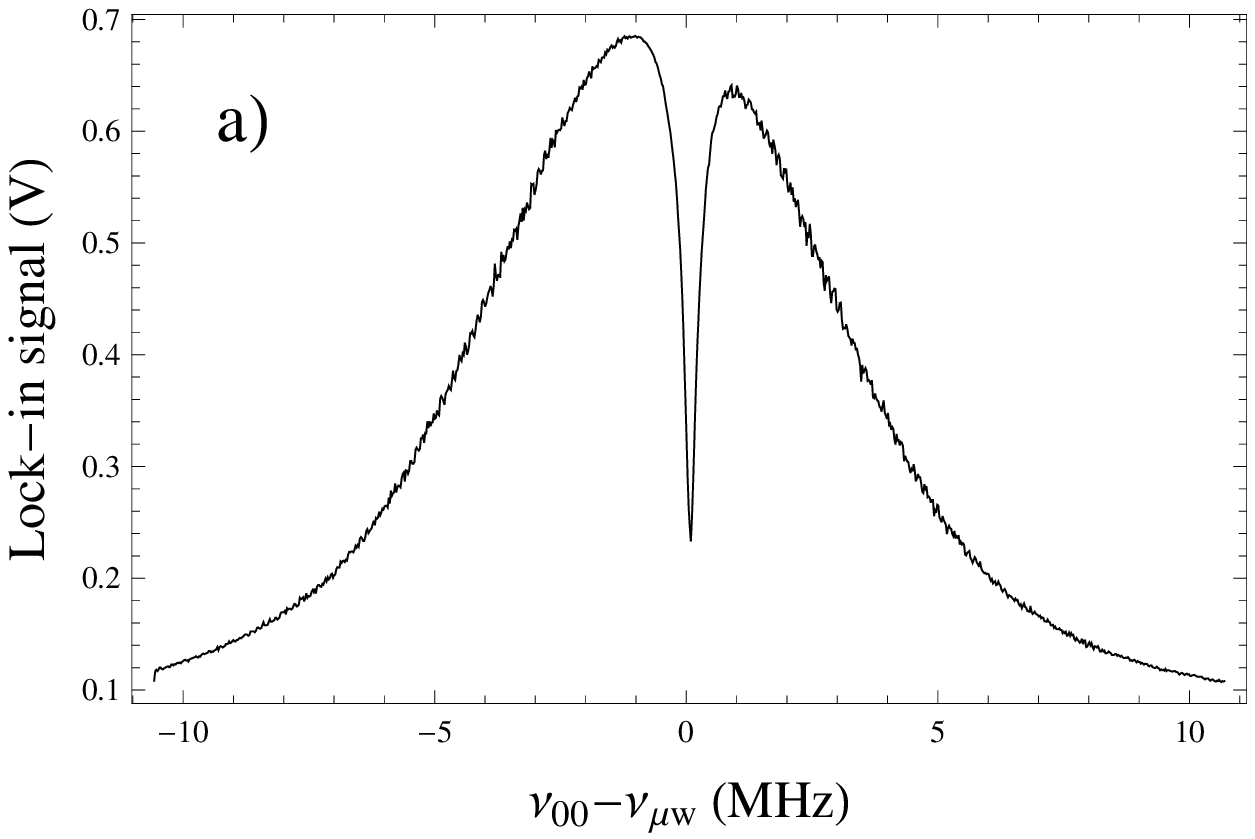}
\includegraphics[width=0.41\textwidth]{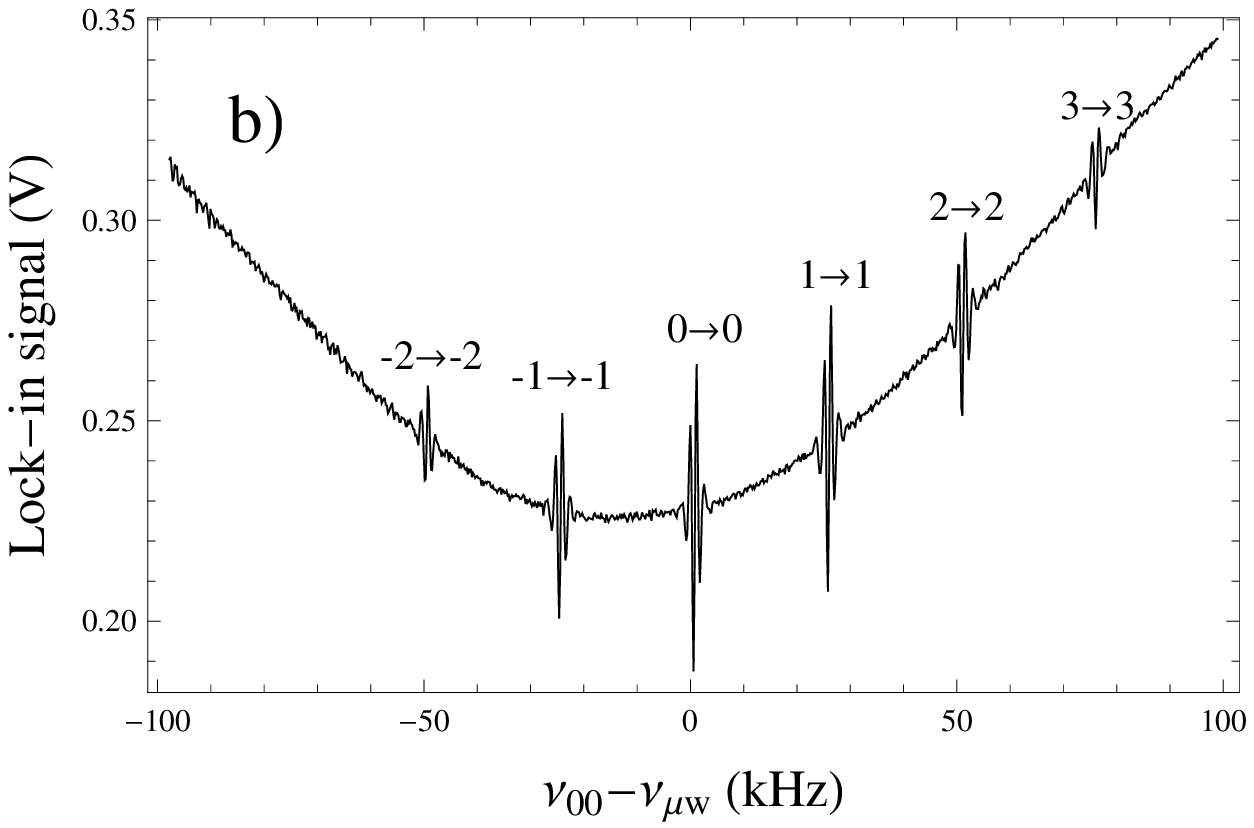}
\label{fig:RamseyLargeScan}
\includegraphics[width=0.41\textwidth]{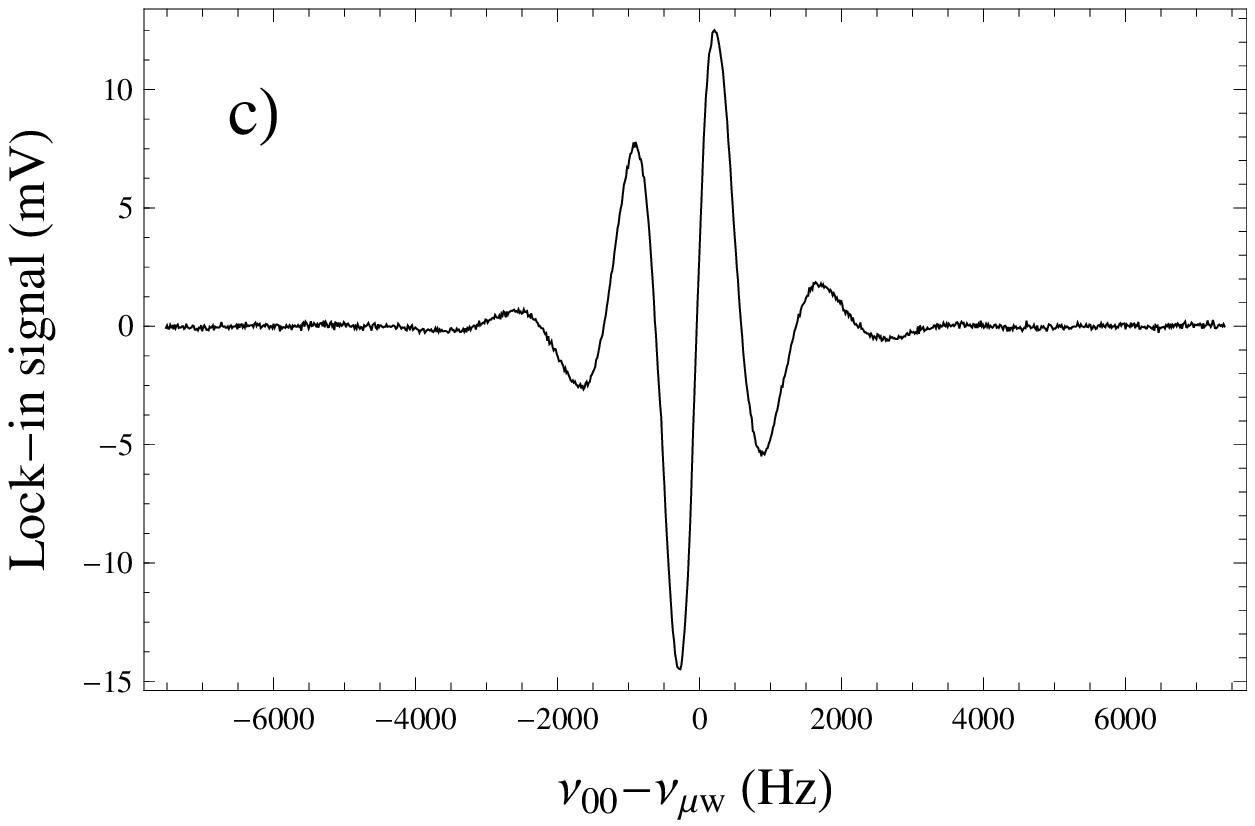}
\caption{Top: Wide microwave frequency scan over the $\ket{6S_{1/2},
F{=}3}{\rightarrow}\ket{6S_{1/2}, F{=}4}$
  Raman transition with central dip due to coherent
  population trapping.
  Center: Zoom into the bottom of the dip revealing
  individual Ramsey resonances of the $\ket{F{=}3, m_{F}}{\rightarrow}
  \ket{F{=}4,m'_{F}{=}m_F}$ transitions, split by the $3.57(1)~\uT$
  magnetic field together with quadratic background. 
  Bottom: Ramsey spectrum of the $\ket{F{=}3, m_{F}{=}0}{\rightarrow}
  \ket{F{=}4, m_{F'}{=}0}$ clock transition after background subtraction.
  In all graphs the microwave frequency $\nurf$ is measured with
  respect to the clock frequency  $\nu_{00}=9.192631770$~GHz. 
 }
\label{fig:RamseyExample}
\end{figure}

Figure~\ref{fig:RamseyExample} shows a large range scan of the
microwave frequency near the $\ket{6S_{1/2}, F{=}3}\rightarrow
\ket{6S_{1/2}, F{=}4}$ Raman transition.
The width of $\sim 9$~MHz reflects the efficiency of the Raman process
that is limited by the (power broadened) width of the excited
$6P_{1/2}$ state.
The ($\sim 400$~kHz wide) dip in the top spectrum is a strongly
power-broadened CPT resonance that occurs in the pump region.
Here the pump beam acts both as pump and probe, similar to microwave
CPT spectroscopy in vapour cells.

The middle graph of Fig.~\ref{fig:RamseyExample} is a zoom into the
bottom of that CPT dip.
It reveals six individual $\Delta m_F{=}0$ Ramsey resonances that are
split by the $3.6~\mu$T magnetic field applied to the beam.
We note that each resonance is superposed on a curved background that
varies from resonance to resonance.
Finally, the lower graph of Fig.~\ref{fig:RamseyExample} shows the
Ramsey pattern of the $\ket{6S_{1/2}, F{=}3, m_F{=}0} \rightarrow
\ket{6S_{1/2}, F{=}4, m_F{=}0}$ clock transition (after background
removal by fitting), whose Stark shift is the object of the
measurement reported below.
All graphs represent the circular dichroism recorded in the probe zone
by the method described in Sec.~\ref{sec:heterodyne} yielding
dispersive fringes since $\Delta\varphi_\mathrm{path}=\pi/2$ for this
measurement.

\subsection{$\mathcal{E}$-field induced frequency and phase shift}
%
As discussed in Sec.~\ref{sec:FieldEffects}, a static electric field
will change both the frequency and the phase of the Ramsey fringes.
The polarizability of interest, defined by Eqs.~(\ref{eq:StarkEqu},
\ref{eq:StarkEqu2}), is inferred from the electric field induced
frequency shift of the Ramsey fringe pattern
\eqref{eq:RamseyFormula} via
\begin{align}
\alpha(m_F)&=-2\,\frac{\Delta\nu_\mathrm{Stark}(m_F)}{\langle|\Efield|^2\rangle_L}
=-2\,\frac{\Delta\nu_\mathrm{Stark}(m_F)}{U^2}\,(d^2)_{\mathrm{eff}}\,,
\label{eq:alphaFromStark}
\end{align}
where $\Delta\nu_\mathrm{Stark}$ is obtained from difference of the
central fringe's center frequency when measured with and without
applied electric field.
The phase shift of the fringes due to the motional magnetic field is
used to test the field modelling predictions.
The following sections describe the results obtained by the two
methods introduced in Sec.~\ref{sec:heterodyne} for determining that
center frequency.

%
\subsection{Stark shifts from Fourier transform analysis}

In order to apply the Fourier transformation described in
Sec.~\ref{sec:FourierTheory}, we have recorded Ramsey spectra, such as the
one shown in the bottom graph of Fig.~\ref{fig:RamseyExample}, for
different voltage differences $U$ applied to the electric field
plates.
Spectra were taken by scanning the microwave frequency over a
typical span of $15~\kHz$, in steps of $15~\Hz$ with $20~\ms$ dwell
time at each frequency value.
The output of the lock-in amplifier measuring the circular dichroism
seen by the probe beam (cf.~Sec.~\ref{sec:heterodyne}) was recorded
on a digital oscilloscope, and typically $24$ complete scans over
each resonance were averaged to yield one Ramsey spectrum as a
function of the microwave frequency.

\begin{figure}[tb]
\centering
\includegraphics[width=0.9\linewidth]{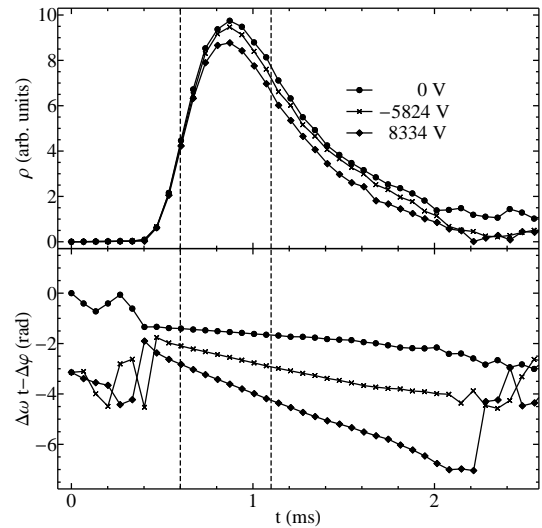}
\caption{Time of flight distribution $\rho(t)$, top, and phase
  $\Delta\omega\,t-\Delta\varphi$, bottom, obtained by Fourier
  transformation of the Ramsey signals under different electric field
  conditions.
  Data between the vertical dashed lines were used to infer the
  frequency, $\Delta\omega(U)$, and phase $\Delta\varphi(U)$ shifts of
  interest. 
} 
\label{fig:TOFandPhaseExamples}
\end{figure}

\begin{figure}[tb]
\centering
\includegraphics[width=0.9\linewidth]{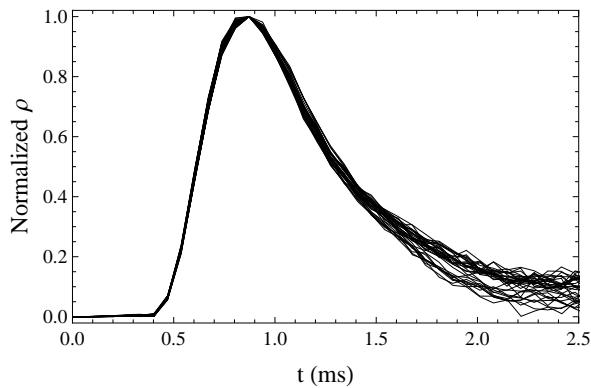}
\caption{Normalized time of flight distributions for all Ramsey
  spectra of the $m_{F}{=}0 \rightarrow m_{F'}{=}0$ transitions
  recorded with different applied voltages.
  For times above $\sim1.6~\ms$ the electric field affects the
  relative atom density.}
\label{fig:TOFm0allNorm}
\end{figure}

The curved background (seen in the middle graph of
Fig.~\ref{fig:RamseyExample}) was removed by fitting the left and
right side of each fringe pattern with a second order polynomial,
and subtracting the fit result from the whole spectrum, yielding the
signal $S(\wrf,U)$.
Following each two measurements with applied voltage, a reference
spectrum $S(\wrf,U{=}0)$ with no applied voltage was recorded.

We next performed numerical Fourier transforms (FT)
of the spectra $S(\wrf, U)$ and $S(\wrf,0)$ recorded with and
without applied voltage to yield the time-of-flight distribution
$\rho(t)$ and the phase $\Delta\omega\,t-\Delta\varphi$. 
A typical result is shown in Fig.~\ref{fig:TOFandPhaseExamples}.
The data show that the phase obeys a linear time dependence only for
a restricted range of time-of-flights.
The signal becomes very noisy
 for very fast and very slow atoms, whose density $\rho(t)$ is small.
While all time of flight distributions are mutually consistent below
$t\sim$1.5~$\ms$, the distribution shows a background for $t>1.5~\ms$
that increases with the electric field intensity.
This systematic dependence on the applied voltage can be seen in
Fig.~\ref{fig:TOFm0allNorm}, where we have superposed
(peak-normalized) time-of-flight distributions recorded with
different applied fields for the $m_{F}{=}0 \rightarrow m_{F'}{=}0$
transition.
Although several interpretations of this effect (including field
dependent forces on the atoms in the field entrance and exit regions
due to the electric field gradients) were attempted, we could find
no model explaining this feature in a quantitative way.
The hyperfine coherence of slow atoms thus seems to acquire a field
dependent phase shift of unexplained origin.

\begin{figure}[tb]
\centering\includegraphics[width=0.45\textwidth]{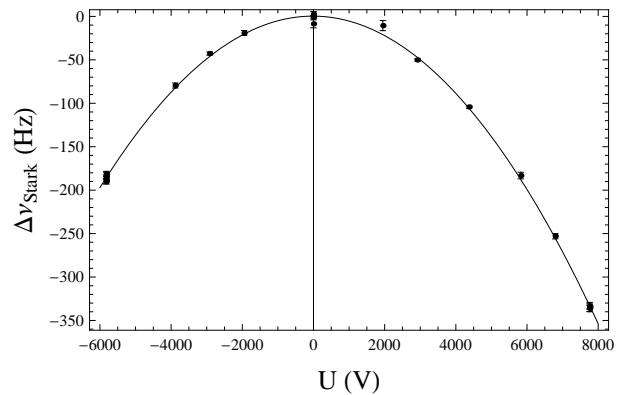}
\caption{Quadratic Stark shift of the $(F{=}3,m_{F}{=}0 \rightarrow
  F{=}4,m_{F}{=}0$ clock transition frequency determined by the Fourier analysis method.
  The solid line is a fit with a second order polynomial. }
\label{fig:StarkShiftVSvolt00}
\end{figure}

Because of the above we have restricted the time range used to fit the
data to $0.6~\ms\le t\le 1.1~\ms$, where both $\rho(t)$ has
significant density, and the phase is well represented by
Eq.~(\ref{equ:fourierphase}), independent of the applied field.
Our error analysis is based on variations of those fit limits.
The phase data in the chosen time interval were fitted by a linear
time dependence, and the electric field-induced frequency and phase
shifts obtained from the fitted slope and intercept according to
\begin{equation}
\Delta\nu_\mathrm{Stark}{=}\Delta\nu(U){-}\Delta\nu(0)
\;\text{and}\;
\Delta\varphi_\mathrm{mot}{=}\Delta\varphi(U)-\Delta\varphi(0)\,.
\end{equation}
The Stark shifts $\Delta\nu_\mathrm{Stark}$ of the clock frequency
extracted in this way are presented in
Fig.~\ref{fig:StarkShiftVSvolt00} together with their fit by a second
order polynomial of the form
$\Delta\nu_\mathrm{Stark}=c_{0}+c_{1}U+c_{2}U^{2}$.
The fitted coefficients $c_{0}= 0.3(9)~\Hz$ and
$c_{1}=-180(110)~\mHz/\kVolt$ are consistent with zero, as expected.

The scalar polarizability $\tspFfour$ is related to the parameter
$c_2$ by
\begin{align}
\alpha_0^{(3)}=-\frac{7}{8}\,c_2\,(d^2)_\mathrm{eff}-\frac{1}{4}\,\alpha_2^{(3)}\,.
\end{align}
Using the fitted value $c_{2}{=}{-}5.51(3)_\mathrm{stat}~\HzkVsq$ and
the literature-based
\cite{Ospelkaus:2003:MFT,Gould:1969:QSS,Carrico:1968:ABR} weighted
average ${-}3.51(16){\times}10^{-2}~\HzkVcms$ for the tensor
polarizability $\ttpFfour$, we find
\begin{subequations}\label{eq:FourierAlpha30}
\begin{align}
\tsp &=2.020(10)_{stat}(9)_{syst}~\HzkVcms \label{eq:FourierAlpha30a}\\
&=2.020(13)~\HzkVcms \label{eq:FourierAlpha30b}\,,
\end{align}
\end{subequations}
and
\begin{subequations} \label{eq:Fourierks}
\begin{align}
\ks&=-2.308\;(11)_{stat}(10)_{syst}~\HzkVcms \label{eq:Fourierksa}\\
&=-2.308\;(15)~\HzkVcms \label{eq:Fourierksb}\,,
\end{align}
\end{subequations}
where we have added the statistical and systematic uncertainties in
Eqs.~\eqref{eq:FourierAlpha30a} and \eqref{eq:Fourierksa} quadratically to
yield the global errors in~\eqref{eq:FourierAlpha30b}
and~\eqref{eq:Fourierksb}, respectively.

The systematic uncertainty is dominated by the $5{\times}10^{-3}$
precision of field calibration constant $(d^2)_\mathrm{eff}$, to
which adds an uncertainty of the pump-probe separation $L$.
The distance between the centers of the pump and probe beams'
intensity distributions was measured to be $L{=}301.1(1)~\mm$.
To correct for the fact that optical pumping occurs with a higher
probability in the upstream part of the atomic-laser beams'
intersection volume, $L$ was lengthened by $\delta L$, taken as half
of the $1/e^{2}$ width of the pump laser beam, and the uncertainty
on $L$ was increased accordingly.
We therefore assign the value of  $L+\delta L=301.8(7)~\mm$ to the
pump-probe separation.
When added quadratically to the uncertainty of $(d^2)_\mathrm{eff}$
the $2{\times}10^{-3}$ effect due to $\delta L$ leads to a
systematic error of $\sim0.6\%$.

Following Eq.~\eqref{eq:StarkEqu}, the relative orientation $\theta$
of the (nominally orthogonal) electric and magnetic fields affects
the contribution from $\ttp$ to $\alpha(0)$.
A mismatch of $\Delta\theta=5\degree$ from perfect orthogonality
yields a relative change of only $10^{-4}$ in the total Stark shift.
By construction, the orthogonality of the two fields is obeyed at
the $\pm 0.5\degree$ level, thus giving a negligible contribution
the systematic error budget.

\begin{figure}[b]
\centering\includegraphics[width=0.45\textwidth]{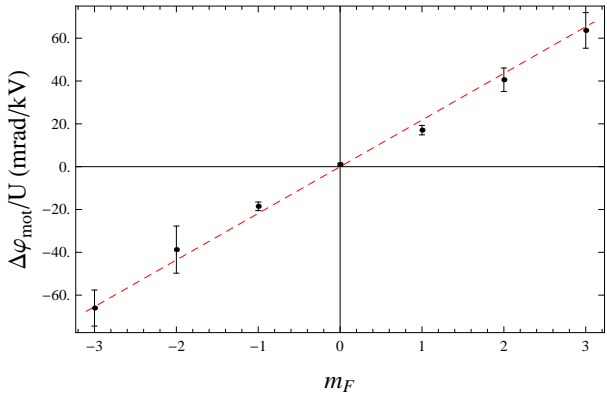}
\caption{Ramsey
  signal phase as function of $m_F$ for the seven possible $\Delta
  m_{F}=0$ hyperfine transitions (each point is the result of a linear
  fit of the phase $\Delta\varphi_\mathrm{mot}(U)$ given by
  Eq.~\eqref{eq:MotionalPhaseShift} as function of the applied voltage
  $U$).
} \label{fig:RamseyPhaseAll}
\end{figure}

We have also measured the Stark shifts of the six $m_{F}{\neq}0$
transitions.
Their Fourier analysis revealed some $m_F$-dependent systematic phase
perturbations that did not allow for an improved determination of the
clock transition's Stark shift.
However, when extrapolated to $t{\rightarrow}0$, the values of the
seven phase plots (equivalent to the one shown for $m_F=0$ in the
bottom graph of Fig.~\ref{fig:TOFandPhaseExamples}) allowed the
extraction of the phase $\Delta\varphi$ in \eqref{equ:fourierphase}.
As discussed in Sec.~\ref{sec:FieldEffects}, one expects a linear
electric field dependent contribution
$\Delta\varphi_\mathrm{mot}\propto\mathcal{E}\propto U$ due to the
motional Zeeman effect.
Being a Zeeman shift, one further expects $\Delta\varphi_\mathrm{mot}$
to have a linear dependence on $m_F$.
Figure~\ref{fig:RamseyPhaseAll} illustrates the anticipated linear
dependence of $\Delta\varphi_\mathrm{mot}$ on $U$ and $m_F$ for all
seven $\Delta m_F=0$ transitions.
The average slope of the fitted line in the figure is
$20(1)~\mu$rad~\Volt$^{-1}m_{F}^{-1}$, while
Eq.~\eqref{eq:MotionalPhaseShift_b}, together with the field integral
of Eq.~\eqref{eq:I1Def} and the modelled numerical value of
Table~\ref{tab:EfieldCallib2} yields
$22(1)~\mu$rad~\Volt$^{-1}m_{F}^{-1}$, thus giving confidence in the
field modelling calculations.

\subsection{Stark shifts from microwave tracking of the resonance frequency}
\begin{figure}[b]
\centering\includegraphics[width=0.45\textwidth]{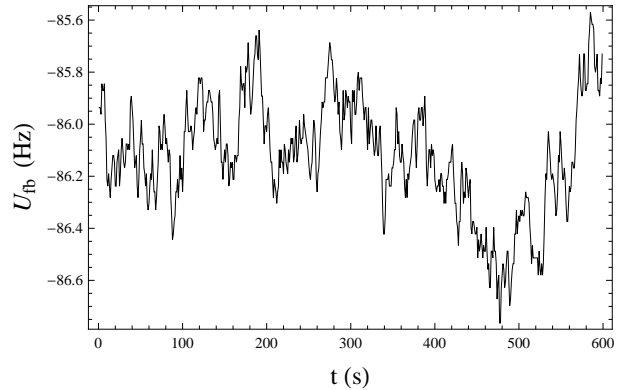}
\caption{Typical feedback signal $U_\mathrm{fb}$ (calibrated in
frequency units) recorded with the microwave frequency locked to the
Ramsey fringe center with a voltage of $-4.066$~kV/cm. The origin of
the ordinate is the frequency generated by the microwave generator,
corresponding here to $9.192631770~\GHz$.} 
\label{fig:UerrHz}
\end{figure}

%
Unfortunately, Stark shift measurements using the frequency-tracking
method (Fig.~\ref{fig:exp_scheme_heterodyne}) described in
Sec.~\ref{sec:heterodyne} were made prior to calibrating the
HV-voltmeter.
A feedback loop controlling the synthesized microwave frequency
$\nurf$ was used to lock the latter to the zero crossing of the
dispersive central Ramsey fringe (Fig.~\ref{FringesPHandFREQchanges}).
For each voltage applied to the electric field generating electrodes
the feedback signal was recorded on a digital oscilloscope for 600~s.
Figure~\ref{fig:UerrHz} shows an example of such a time series.

\begin{figure}[ht]
\centering
\includegraphics[width=0.9\linewidth]{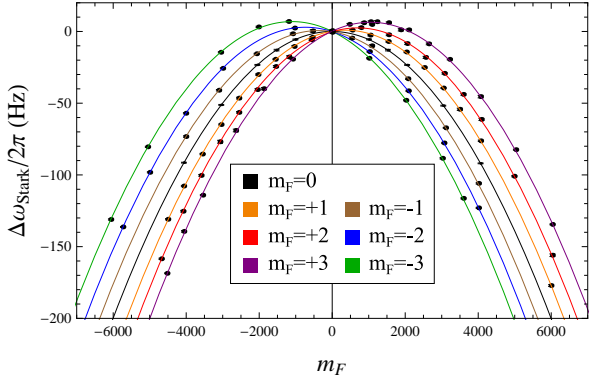}
\caption{(Color Online) Stark frequency shift of all seven $\Delta
  m_{F}=0$
  hyperfine transitions in the $\Efield \perp \vec{\Bfield}$
  configuration with $\sigma^{+}$--$\sigma^{+}$ polarized optical fields.
  The lowest $m_{F}$ value corresponds to the left-most curve and $m_{F}$ 
  is increasing from left to right.
  The fitting function are polynomials of second order in $U$.
  The lateral displacement of the parabolas is due to the motional
  magnetic field, while their curvature results from the Stark
  interaction proper.} 
  \label{fig:StarkShiftAll_LTCF}
\end{figure}

\begin{figure}[b]
\centering
\includegraphics[width=0.9\linewidth]{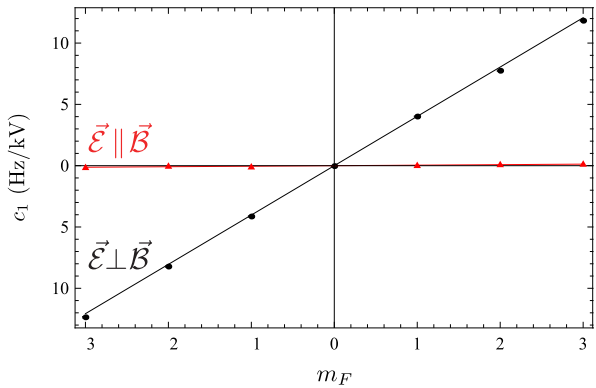}
\caption{(Color Online) $m_F$-dependence of the fit coefficients $c_1$
  representing the linear (motional field induced) Stark shift of the
  $F{=}3,m_F\rightarrow F{=}4,m_F$ transitions.
The (red) triangles show data from Table~\ref{tab:LTFfitvaluesparra})
for the $\Efield{\parallel}\vec{\Bfield}$ configuration, (black) dots
(data from Table ~\ref{tab:LTFfitvaluesperp}) for the
$\Efield{\perp}\vec{\Bfield}$ configuration.
The lines are linear fits to the data. As expected, the motional
magnetic field shift vanishes in the  $\Efield \parallel
\vec{\Bfield}$ configuration.} 
\label{fig:LTFpolynomialcoef}
\end{figure}

\newcolumntype{e}[1]{D{.}{}{#1}}
\begin{table}[ht]
\caption{Polynomial fit parameters for the $\Efield{\perp}
\vec{\Bfield}$ configuration.}
\begin{tabular}{c  k{6}  k{7} k{8} } \hline\hline
$m_{F} \rightarrow m_{F'}$\datastrut
                   &\multicolumn{1}{c}{$c_{0}~(\Hz)$}
                              &  \multicolumn{1}{c}{$c_{1}~(\HzperkVolt)$}
                                         &\multicolumn{1}{c}{$c_{2}~[\HzperkVsquared]$}\\
\hline
$\phantom{+}0\rightarrow \phantom{+}0$
                   & 0.01(9)  &-0.036(21)& 5.532(4) \\
$+1\rightarrow +1$ & 0.17(14) & 4.01(4)  & 5.593(14)\\
$-1\rightarrow -1$ &-0.16(24) &-4.14(5)  & 5.573(27)\\
$+2\rightarrow +2$ &-0.19(20) & 7.74(4)  & 5.588(15)\\
$-2\rightarrow -2$ &-0.09(18) &-8.23(5)  & 5.580(14)\\
$+3\rightarrow +3$ & 0.22(15) & 11.83(3) & 5.645(9) \\
$-3\rightarrow -3$ & 0.06(18) &-12.37(5) & 5.601(15)\\
\hline\hline
\end{tabular}
\label{tab:LTFfitvaluesperp}
\caption{Polynomial fit parameters for the $\Efield{\parallel}
\vec{\Bfield}$ configuration. The  $m_{F}{=}0 \rightarrow
  m_{F'}{=}0$ clock transition is forbidden in this configuration.
}
\begin{tabular}{c  k{6}  k{7} k{8} } \hline\hline
$m_{F} \rightarrow m_{F'}$\datastrut
                   &\multicolumn{1}{c}{$c_{0}~(\Hz)$}
                              &  \multicolumn{1}{c}{$c_{1}~(\HzperkVolt)$}
                                        &\multicolumn{1}{c}{$c_{2}~[\HzperkVsquared]$}\\
\hline
 $\phantom{+}0\rightarrow \phantom{+}0$
                  &\multicolumn{1}{c}{---}
                              &\multicolumn{1}{c}{---}
                                         & \multicolumn{1}{c}{---}\\
$+1\rightarrow +1$&  0.55(45) &-0.011(6)  & 5.609(17)\\
$-1\rightarrow -1$&  1.09(37) &-0.120(50) & 5.605(8) \\
$+2\rightarrow +2$&  0.02(40) & 0.068(47) & 5.577(14)\\
$-2\rightarrow -2$& -0.11(42) &-0.061(57) & 5.573(17)\\
$+3\rightarrow +3$& -0.08(52) & 0.110(70) & 5.532(23)\\
$-3\rightarrow -3$& -0.46(52) &-0.160(60) & 5.544(24)\\ \hline\hline
\end{tabular}
\label{tab:LTFfitvaluesparra}
\end{table}

Stark shift measurements were made for two different relative
orientations of the electric and magnetic fields, viz.,
$\Efield{\perp}\vec{\Bfield}$, corresponding to
$f(\theta{=}\pi/2) = -1$, and $\Efield {\parallel}\vec{\Bfield}$,
equivalent to $f(\theta{=}0) = 2$.
Measurements with the former configuration used
$\sigma^{+}$--$\sigma^{+}$ polarized components of the bichromatic
optical field, while their polarization was $\pi$--$\pi$ (with respect
to $\vec{\Bfield}$) in the latter configuration.
The dependence of the central fringe's shift on the applied electrode
voltage $U$ was recorded for all individual $\Delta m_F=0$ hyperfine
transitions in both field configurations.
The data was fitted with a second order polynomial of the form
$\Delta\omega_\mathrm{Stark}/2\pi=c_{0}+c_{1}U+c_{2}U^{2}$.
Figure~\ref{fig:StarkShiftAll_LTCF} presents data and the parabolic
fit curves for all hyperfine transitions in the
$\Efield{\perp}\vec{\Bfield}$ configuration.
The $m_F$-dependence of the fit parameters $c_{1}$ for both
configurations is shown in Fig.~\ref{fig:LTFpolynomialcoef} and
numerical values of all parameters $c_{i}$ are given in
Tables~\ref{tab:LTFfitvaluesperp} and~\ref{tab:LTFfitvaluesparra}.
Note that the $m_{F}{=}0 \rightarrow m_{F'}{=}0$ transition is
forbidden in the $\Efield \parallel \vec{\Bfield}$ configuration.

No physical significance can be attributed to the constant
$c_{0}$, which is indeed consistent with zero.
The linear coefficient $c_{1}$ is due to the motional magnetic field
induced phase shift $\Delta\varphi_{\mathrm{mot}}$ of
Eqs.~\eqref{eq:MotionalPhaseShift}, which influences the
$\Delta\omega_\mathrm{Stark}(U)$ dependence in different ways for the
two field configurations.
In the $\Efield{\parallel}\vec{\Bfield}$ configuration, the motional
magnetic field ($\approx$nT) is perpendicular to the applied magnetic
field of $\sim$4~$\mu$T, so that the motional field adds quadratically
to the latter, thus giving a negligible contribution.
As expected, the fitted $c_1$ coefficients for $\Efield{\parallel}
\vec{\Bfield}$ are compatible with zero within three standard
deviations.
In the $\Efield{\perp}\vec{\Bfield}$ configuration, on the other hand,
the motional magnetic field is directed along the applied magnetic
field, and thus increases/decreases the latter directly.
Being a Zeeman effect, the motional field effect is expected to be
proportional to $m_F$, a feature that is well obeyed by the
experimental data shown in Fig.~\ref{fig:LTFpolynomialcoef}.

The fitted $c_2$ coefficients  are related to the Stark
polarizabilities $\alpha(m_F)$ and the polarizability of interest
\tsp{} via
\begin{subequations}\label{eq:alphFringeLock}
\begin{align}
    \frac{\alpha(m_F)}{2}&=-\,c_2\,(d^2)_\mathrm{eff}\label{eq:alphac2}\\
    &=
    \left(\frac{8}{7}\,\tsp+\frac{4}{7}\,f(\theta)\,\ttp\right)-\frac{3}{28}\,f(\theta)\,\ttp\,m_F^2\label{eq:alphaalpha02}\,.
\end{align}
\end{subequations}
where $\alpha_\mathrm{k}^{(3)}=\alpha_\mathrm{k}^{(3)}(F{=}4)$, and
where $f(\theta) = +2$ and $-1$ for the
$\Efield{\parallel}\vec{\Bfield}$ and $\Efield{\perp}\vec{\Bfield}$
configurations, respectively.
The polarizability
$\alpha(m_F)=-2\,\Delta\nu_\mathrm{Stark}(m_F)/\mathcal{E}^2$,
extracted from the measured Stark shifts
$\Delta\nu_\mathrm{Stark}(m_F)$ using
$\mathcal{E}^2=U^2/(d^2)_\mathrm{eff}$ has $m_{F}$-independent, and
$m_{F}$-dependent contributions from the third order polarizabilities
\tsp{} and \ttp{}, with \ttp{} being $\sim$2 orders of magnitude
smaller than \tsp{}.

\begin{figure}[ht]
\includegraphics[width=0.9\linewidth]{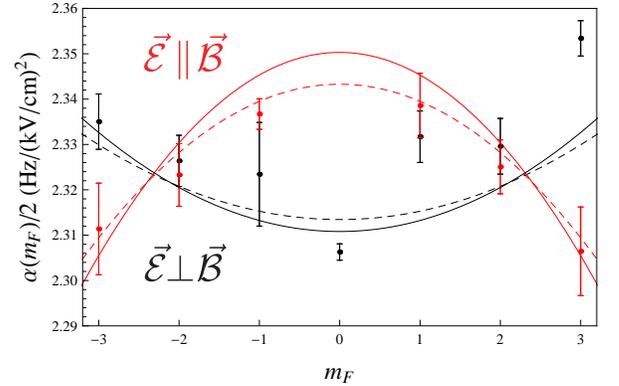}
\caption{(Color Online)
$m_F$-dependence of the differential Stark polarizability
  $\alpha(m_F)/2$ from Ramsey fringe frequency tracking experiments
  using two field configurations, together with parabolic fits.
  The solid lines are fits based on Eq.~\eqref{eq:alphFringeLock} with
  both $\tsp$ and $\ttp$ as free parameters.
  The dashed lines are fits with $\ttp$ fixed to the best literature
  value. 
}
\label{fig:StarkShiftPola_LTCF}
\end{figure}

The experimental values of $\alpha(m_F)/2$ inferred from the fitted
coefficients $c_2$ following Eq.~\eqref{eq:alphac2} are shown in
Fig.~\ref{fig:StarkShiftPola_LTCF}, together with fitted curves and
exhibit the anticipated quadratic $m_F$-dependence, with curvatures in
the ratio $(-1):2$ as predicted by the $f(\theta)$-dependence.
Note that based on Eq.~\eqref{eq:alphaalpha02}, the fitted curves are
expected to intersect at the `fictitious' $m_F$ value of
$m_F^\ast=\sqrt{16/3}\approx~2.31$, for which
$\alpha(m_F^\ast)/2 = 8\tspFfour/7 = {-}k_s$.

We have taken two approaches for fitting Eq.~\eqref{eq:alphaalpha02} to
the parabolas of Fig.~\ref{fig:StarkShiftPola_LTCF}.
In a first approach we have left both $\tspFfour$ and $\ttpFfour$ as
free parameters, and in a second approach we have left only
$\tspFfour$ as free parameter, by fixing $\ttpFfour$ to
\begin{equation}
  \ttpFfour=-0.0351(16) \HzkVcms\,,
\label{eq:alpha2Lit}
\end{equation}
taken as the weighted average of published measurements
\cite{Ospelkaus:2003:MFT,Gould:1969:QSS,Carrico:1968:ABR}.
The two fit methods yield the same value for the scalar polarizability
\begin{subequations}\label{eq:alpha0FringeLock}
\begin{align}
  \tspFfour&=2.033(1)_{stat}(22)_{syst}\,\HzkVcms 
  \label{eq:alpha0FringeLockStatSyst}\\
  &=2.033(22)\,\HzkVcms 
  \label{eq:alpha0FringeLockTot}\,,
\end{align}
\end{subequations}
and the scalar Stark shift parameter
\begin{subequations}\label{eq:ksFringeLock}
\begin{align}
  \ks&=-2.323 (1)_{stat}(25)_{syst}\,\HzkVcms\,
  \label{eq:ksFringeLockStatSyst}\\
  &=-2.323(25)\,\HzkVcms
  \label{eq:ksFringeLockTotb}\,,
\end{align}
\end{subequations}
respectively, where the errors in \eqref{eq:alpha0FringeLockTot} and
\eqref{eq:ksFringeLockTotb} represent the squared sums of statistical
and systematic errors, respectively.

When fitted as free parameter, we obtain a tensor polarizability
\begin{equation}\label{eq:alpha2FringeLock}
  \ttpFfour=-0.046(4) \HzkVcms\\
\end{equation}
that agrees with the literature average within two standard
deviations.

The relative statistical uncertainty of $\tspFfour$ obtained by the
frequency-tracking method is rather small $\sim 5 \times 10 ^{-4}$.
However, the systematic uncertainty on the polarizability in the
fringe-tracking experiment is dominated by imprecise knowledge of the
voltage drop across the field electrodes.
In those early experiments the $200~\MOhm$ protection resistor was
in series with the field-producing capacitor, but the voltage was
measured directly at the supply output.
Any leakage current across the field plates will thus lead to a
voltage drop over the protection resistor, thereby lowering the
effective voltage applied to the electrodes.
We estimate the systematic field uncertainty due to this unfortunate
configuration as follows:
Based on the current ($\sim$1$\muAmpere$) drawn from the power supply
and the digital high voltage voltmeter's $6.5~\GOhm$ internal
resistance, we estimate the uncertainty on the electrode voltage $U$
to be $\sim 5 \times 10^{-3}$.
Quadratically adding the latter uncertainty and the
$\sim 5 \times 10^{-3}$ uncertainty of the field calibration
constant $(d^2)_\mathrm{eff}$ yields the systematic errors of
Eqs.~\eqref{eq:alpha0FringeLock} and \eqref{eq:ksFringeLock}.

\section{Summary and conclusion}
\begin{figure}[b]
\centering\includegraphics[width=\columnwidth]{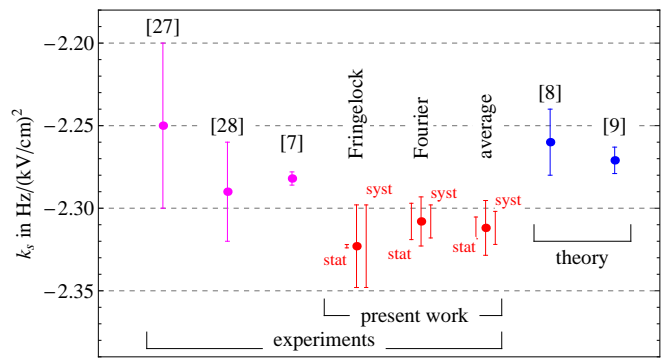}
\caption{(Color Online) Comparison of results from present work with
 previous experimental and theoretical values for the Cs clock
 transition's scalar Stark shift coefficient $\ks$.
The statistical and systematic uncertainty intervals of our results
are shown on the left and right, respectively, of the experimental
data points.
The errors on our data points were derived from those errors as
discussed in the text.
Note that the recent value $\ks=2.050(40)~\HzkVcms$ of
Godone~\etal~\cite{Godone:2005:SSM} is far off scale of the present
plot.}
\label{fig:MeasAlphaFin}
\end{figure}

%
We have used two separate methods to measure the differential third
order electric polarizability of the Cs ground state hyperfine levels,
from which we infer the third order scalar polarizability \tspFfour{}
of the $F{=}4$ state and the scalar Stark shift coefficient $\ks$
which are related by
\begin{equation}
 \ks = -\frac{8}{7}\,\tspFfour\,.
\label{eq:DeltaAlphaDef}
\end{equation}
The result obtained by the fringe tracking method has a very small
statistical error but suffers from a large systematic uncertainty
because of imprecisions in the voltage measurement.
The $\alpha^{(3)}_0(F{=}4)$ value obtained by the Fourier analysis
method, on the other hand, has comparable statistical and systematic
errors.
The individual results and their average value are presented in
Fig.~\ref{fig:MeasAlphaFin}, together with past experimental and
theoretical values.
For a comparison with more previous results we refer the reader to
Fig.~3 of Ref.~\cite{Rosenbusch:2007:BRS}.

The weighted average of our two independent measurements, expressed in
terms of the scalar Stark shift coefficient \ks{}, yields
\begin{align}
\ks   &= -\frac{8}{7}\tspFfour \nonumber\\
      &=  -2.312(7)_{stat}(10)_{syst}~\HzkVcms\,.
      \label{eq:ksfinal0}
\end{align}
The (quadratically) combined statistical and systematic errors of the
individual measurements given in Eqs.~\eqref{eq:Fourierksb}
and~\eqref{eq:ksFringeLockTotb}, respectively, were used as weights to
derive the final value.
The statistical error on the final value is the statistical error of
the mean.
The systematic error of the final value represents the systematic
uncertainty of the Fourier analysis data, the (larger) systematic
error of the fringe tracking data playing a subordinate role because
of the reduced contribution of those data to the final result
\begin{equation}
\ks = -2.312(17)~\HzkVcms\,,
\label{eq:finalksa}
\end{equation}
where we have taken the conservative approach by (linearly) adding the
statistical and systematic errors of~\eqref{eq:ksfinal0}.

The \ks-value of~\eqref{eq:finalksa} is equivalent to
\begin{align}
\tspFfour &=2.023(6)_{stat}(9)_{syst}  \\
&=2.023(15)~\HzkVcms\,,
\end{align}
and 
\begin{align}
\beta &= \frac{\ks}{\nu_{00}}\,
         \left(\!831.9\,\frac{\mathrm{V}}{\mathrm{m}}\right)^2
\label{eq:betaisolation}\\
      &=-1.7406(53)_{stat}(75)_{syst}\times 10^{-14}\,.
\end{align}

Our result has a relative error of 0.7\% and is to be compared to the
most precise experimental value published to
date~\cite{Rosenbusch:2007:BRS}
\begin{align}
\ks   &= 2.282(4)~\HzkVcms\nonumber\,,
\end{align}
with a relative uncertainty below 0.2\% that testifies to the
remarkable control of the field integral in that experiment.
Our result differs by $\sim2$~standard deviations from that
measurement and from theoretical predictions, and can be considered to
be in agreement with those results.

\begin{acknowledgments}
 Work funded by the Swiss National Science Foundation, grants
 \#200020--126499 and \#200021--117841.  Support from the University
 of Fribourg Fonds de Recherche, and the Physics Department's
 mechanical and electronic workshops is acknowledged.
 Dr.~Z.~Andjelic and ABB Corporate Research, Baden,
  Switzerland, are thanked for their generous donation of time and
  resources for the field modelling.
\end{acknowledgments}

\raggedright
\bibliography{FRAPref}

\end{document}